\documentclass{webofc}
\usepackage[varg]{txfonts}   
\newcommand{\beq}{\begin{eqnarray}}
\newcommand{\eeq}{\end{eqnarray}}
\newcommand{\eq}{eqnarray}

\newcommand{\al}{{\alpha}}

\newcommand{\ci}{\cite}

\newcommand{\de}{{\delta}}
\newcommand{\De}{\Delta}

\newcommand{\ka}{{\kappa}}
\newcommand{\la}{{\lambda}}
\newcommand{\La}{{\Lambda}}

\newcommand{\om}{{\omega}}

\newcommand{\pa}{{\partial}}
\newcommand{\no}{{\nonumber}}
\newcommand{\f}{\frac}

\newcommand{\calB}{{\cal B}}

\newcommand{\calG}{{\cal G}}
\newcommand{\calH}{{\cal H}}

\newcommand{\calR}{{\cal R}}

\newcommand{\Ho}{Ho\v{r}ava~}
\newcommand{\cA}{{\cal A}}

\newcommand{\cH}{{\cal H}}

\begin{document}
\title{On Gauge Invariant Cosmological Perturbations
in UV-modified Ho\v{r}ava Gravity: A Brief Introduction\footnote{
Proceeding of   International Joint Conference of\, ICGAC-XIII and IK15 (July 3-7, 2017, Seoul, Republic of Korea)}}
%
%

\author{\firstname{Mu-In} \lastname{Park}\inst{1} \thanks{\email{muinpark@gmail.com}}
}

\institute{Research Institute for Basic Science, Sogang University,
Seoul, 121-742, Korea }

\abstract{
We revisit gauge invariant cosmological perturbations in UV-modified, $z=3$ Ho\v{r}ava gravity with one scalar matter field,
which has been proposed as a renormalizable gravity theory without the
ghost problem in four dimensions. We confirm that there is no extra graviton modes and general relativity is recovered in IR, which achieves the consistency of the model. From the UV-modification terms which break the detailed balance condition in UV, we obtain scale-invariant power spectrums for {\it non}-inflationary backgrounds, like the power-law expansions, without knowing the details of early expansion history of Universe. This could provide a new framework for the Big Bang cosmology.
}
\maketitle

{\it ``All truth passes trough three stages. First, it is ridiculed. Second, it is violently opposed. Third, it is accepted as being self-evident." (A. Schopenhauer)}\\
\section{Introduction}
\label{sec-1}
(1). As a particle physicist's point of view, a surprise of cosmology is as follows. In particle physics, we need (fundamental) scalar fields (called Higgs fields) for ``theoretical" reasons ({\it i.e.}, to give masses to fundamental particles with renormalizability), but we have waited for a long time (about 40 years, after 1964 papers) for an experimental confirmation at LHC (4 July, 2012). This is (thought to be) the first elementary scalar particle discovered in Nature.

However, in cosmology, it has been known for a long time that there {\it is} scalar (cosmological fluctuation mode) in the sky after the discovery of CMB (1964, the same year that Higgs field has been proposed !). But, the existence of a scalar mode in the sky is a big mystery from the following reasons. The small fluctuations in CMB are thought to be due to (space-time)  fluctuations in the early epoch of our (Big Bang) Universe, which should be described by General Relativity (GR). However, GR alone can {\it not} have scalar (fluctuation) mode but only the (usual) tensor mode (2 polarizations, called gravitational waves) ! The cosmological/primordial tensor modes have not been discovered yet, though astrophysical tensor modes (gravitational waves) from black hole mergers have been detected in LIGO. So, it would not be quite strange (at least to me) even though we discover the cosmological/primordial tensor modes (called B-mode) in the near future. Rather, the existence of the scalar mode is much more surprising !

The simplest way to explain the data is introduce a ``cosmic scalar" field
and ``assume" some peculiar behaviors in the early epoch (known as the inflationary epoch). But, there are huge numbers of explicit models and we still do not know what is the right one and its origin--Is Higgs the remnant of the primordial scalar ? This situation is opposite to that of particle physics: In cosmology, we have discovered scalar mode (CMB) long ago but we do not have its theory yet !

Another way to explain the scalar mode is to modify GR, like $f(R)$, massive gravity, etc., so that the gravity itself has additional scalar modes. But, in that case the scalar (gravitation) could be disaster since it could affect the known (or, well-established) GR test in solar system (,{\it i.e.,} low energy (IR) test of GR), unless we decouple it from low energy GR sector. And also, this could affect the dark matter and dark energy problems as well. Sometimes, the scalar could be ghost as well, which should be avoided. Today, I will consider another modified gravity theory but without the scalar gravitation mode so that the usual scalar matters are needed in explaining the cosmology data.

(2). One the other hand, our universe is considered to be created from ``quantum
(vacuum) fluctuations". Actually, since the scalar power spectrum in inflationary cosmology contains Planck constant $\hbar$ as well as Newton's constant $G$ ($H$ is the Hubble parameter),
$
\Delta^2(k)= 8 \pi G \hbar  ( {2 H^2}/{\pi^2}),
$
one can consider the power spectrum as a ``quantum gravity effect" and so does the inflationary cosmology as a ``quantum cosmology" ! Moreover, recent LIGO's detections of gravitational waves from merging black holes seems to imply that we need to consider quantum gravity more seriously. Actually, we open the strong gravity test era of GR, beyond the weak gravity test in solar system.

(3). Do we have quantum gravity then ? For quantum theory of particle's interactions, ``renormalization" has been a powerful constraint and Higgs particle is its natural consequence. Now, what if we require renormalizability in quantum gravity ? But, it is well-known that the renormalizable quantum gravity can not be realized in Einstein's gravity or its (relativistic) higher-derivative generalizations: There are ``ghosts" which have kinetic terms with a wrong sign, in addition to massless gravitons.

Recently, Ho\v{r}ava (or Ho\v{r}ava-Lifshitz (HL)) gravity has been proposed as a renormalzable gravity \ci{Hora}. There is no ghosts (in the tensor modes) by abandoning the equal-footing treatment of space and time ({\it i.e.,} Lorentz symmetry) in the higher energy regime (UV). It is power-counting renormalizable but no proof of renormalizability yet. This resembles the situations of Yang-Mills theory when it first appeared.

In cosmology, this theory {\it may} provide inflationary effect without ``inflationary phase ({\it i.e.}, early de Sitter or accelerating phase)" so that scale-invariant (scalar/tensor) spectrum can be also produced. This has been first argued by Mukohyama \cite{Muko:1007} but there is no rigorous analysis about this yet. On the contrary, in the works of Gao {\it et. al.} \ci{Gao} and Gong {\it et. al.} \ci{Gong}, scalar spectrum are not scale-invariant though tensor spectrum does. This is today's topic and it provides a natural formulation of cosmology, known as ``effective field theory (EFT)", by construction.

\section{Ho\v{r}ava (-Lifshitz) Gravity: Basic Idea}

In 2009, Ho\v{r}ava proposed a renormalizable, higher-derivative gravity
theory, without ghost problems, by abandoning Lorentz symmetry in UV but
keeping, so called, ``foliation preserving" diffemorphism ($FPDiff$) \ci{Hora}.

The basic ingredients of the Ho\v{r}ava are (1) Einstein gravity (with a
Lorentz deformation parameter $\lambda$), (2) non-covariant deformations
with higher-spatial derivatives (up to six orders), and (3)
``detailed balance" in the coefficients (with five constant parameters,
$\kappa, \la, \nu, \mu, \La_W$), in contrast to the Einstein gravity
case with the three physical parameters $c, G, \La$ and $\la=1$.
Here, six orders of spatial derivatives $(z=3)$ came from the power counting renormalizability in $3+1$ dimensions. (For the limitation of space and time, I can not explain all the details of the construction. So, please read the original work of Ho\v{r}ava \ci{Hora} about this.)

From the Ho\v{r}ava's construction, some improved UV behaviors, without ghosts, are expected so that the gravity theory ``could" be renormalizable and one might have predictable quantum gravity theory. But, in cosmology applications, it seems that the detailed-balance condition is too strong to get the required, scale invariant cosmological perturbations. For example, tensor spectrum is scale invariant but scalar spectrum is not ! \cite{Gao, Gong}. So, we need to break the detailed balancing in some way but without altering the improved UV behaviors of scale invariant tensor modes in the previous works \cite{Gao, Gong} and we called it as ``UV modifications" \ci{Shin}.

\section{Cosmological Perturbations in Ho\v{r}ava gravity}

We start by considering the four-dimensional, UV-modified \Ho
gravity action with $z=3$, which is power-counting renormalizable \ci{Hora},
\begin{\eq}
\label{HL action}
S_\mathrm{g} &=& \int d \eta d^3x \sqrt{g} N \left[ \frac{2}{\kappa^2}
\left(K_{ij}K^{ij} - \lambda K^2 \right) - {\cal V} \right], \\
-{\cal V}&=& \sigma+ \xi R + \alpha_1 R^2+ \alpha_2 R_{ij}R^{ij}
+\alpha_3 \frac{\epsilon^{ijk}}{\sqrt{g}}R_{il}\nabla_jR^l{}_k
 + \alpha_4 \nabla_{i}R_{jk} \nabla^{i}{R}^{jk}
+\al_5 \nabla_{i}R_{jk}\nabla^{j} {R}^{ik}
+\al_6 \nabla_{i}R\nabla^{i}R  
, \no \label{V_Horava}
\end{\eq}
where (the prime $(')$ denotes the derivative with
respect to $\eta$)
$
K_{ij}=\frac{1}{2N}\left({g_{ij}}'-\nabla_i N_j-\nabla_jN_i\right)\, ,~
ds^2 =-N^2 c^2 d\eta^2 +g_{ij}\left(dx^i+N^i d\eta\right) \left(dx^j+N^j d\eta\right). \no
$
With the detailed balance condition, the number of
independent coupling constants can be reduced to six, {\it i.e.},
$\kappa,\lambda,\mu,\nu,\La_W,\om$ for a viable model in IR \ci{Keha,Park:0910},
%
$
\sigma = \f{3 \kappa^2 \mu^2 \La_W^2}{8 (3 \la-1)},~
\xi=\f{\ka^2 \mu^2 (\omega-\La_W)}{8 (3 \la-1)},~
\al_1=\f{\ka^2 \mu^2 (4 \la-1)}{32 (3 \la-1)}, ~
\al_2=-\f{\ka^2 \mu^2 }{8}, ~
\al_3 = \f{\ka^2 \mu }{2 \nu^2},~\al_4=-\f{\ka^2}{8 2 \nu^4}=-\al_5=-8 \al_6,
$
%
But, we do not restrict to this case only, at least for the
UV couplings $\al_4,\al_5,\al_6$ so that the power-counting renormalizable
and scale-invariant cosmological scalar fluctuations can be obtained.

For the power-counting renormalizable matter action, we consider $z=3$ scalar
field action \ci{Calc,Kiri},
\begin{equation}
\label{matter action}
S_\mathrm{m} = \int d\eta d^3x \sqrt{g} N \left[ \frac{1}{2N^2}
 \left( \phi' - N^i \pa_i \phi\right)^2 - V(\phi)- Z(\pa_i\phi)  \right] \, ,
\end{equation}
where
%
$
Z(\pa_i\phi) = \sum_{n=1}^3 \xi_n \partial_i^{(n)}\phi \partial^{i(n)}\phi \,
$
%
with the superscript $(n)$ denoting $n$-th spatial derivatives, and
$V(\phi)$ is the matter's potential without derivatives.

The actions, $S_\mathrm{g}$ and $S_\mathrm{m}$ are invariant under
the foliation preserving {\it FPDiff} \ci{Hora},
\begin{\eq}
\label{Diff}
\delta x^i &=&-\zeta^i (\eta, {\bf x}), ~\delta \eta=-f(\eta),~
 \delta
g_{ij}=\pa_i\zeta^k g_{jk}+\pa_j \zeta^k g_{ik}+\zeta^k
\pa_k g_{ij}+f g'_{ij},\\
\delta N_i &=& \pa_i \zeta^j N_j+\zeta^j \pa_j N_i+\zeta^{'j}
g_{ij}+f {N'}_i+f' N_i,~
\delta N= \zeta^j \pa_j N+f N'+f' N, ~
\delta \phi=\zeta^j \pa_j \phi +f \phi'. \no
\end{\eq}

In order to study the cosmological perturbations around the homogeneous and isotropic backgrounds (as seen in CMB), we expand the metric and the scalar field as,
\begin{\eq}
\label{pert}
N =  a(\eta)[1+{\cal A}(\eta,{\bf x})] \, ,~
N_i =  a^2(\eta){{\cal B}(\eta,{\bf x})}_i \, , ~
g_{ij} =  a^2(\eta) [\de_{ij}+h_{ij} (\eta,{\bf x})],~
\phi = \phi_0(\eta) + \delta\phi(\eta,{\bf x}) \, ,
\end{\eq}
by considering spatially flat ($k=0$) backgrounds and the conformal (or {\it comoving}) time $\eta$, for simplicity. By substituting the metric and scalar
field of (\ref{pert}) into the actions one can obtain the linear-order
perturbation part of the total action $S=S_g +S_m$
which gives the Friedman's equations for the background,
\begin{\eq}
\label{friedmann_eq}
 &&\calH^2 =
-\frac{\kappa^2}{6(1-3\lambda)} \left( \f{1}{2}{\phi_0'}^2
 + a^2 \left(V_0 -\sigma \right) \right) \, ,~
  \calH^2 +2\calH' =
\frac{\kappa^2}{2(1-3\lambda)} \left( \f{1}{2}{\phi_0'}^2
 - a^2 \left(V_0 -\sigma \right) \right) \, , \no
\\
\label{phi_evolution}
&&\phi_0'' + 2\calH\phi_0' + a^2V_{\phi_0} = 0 \, ,
\end{\eq}
with the comoving Hubble parameter ${\cal H} \equiv a'/a$, $V_0 \equiv V(\phi_0), V_{\phi_0}\equiv (\pa V/\pa \phi)_{\phi_0}$, and $h\equiv h^i_i$.
Here, it important to note that there is {\it no} higher-derivative corrections to the Friedman's equations for spatially flat case and so the background equations are the same as those of GR \ci{Keha,Park:0910}. However, even in this case, the higher-derivative effects can {\it re}appear in the perturbed parts.

The quadratic part of the total perturbed action is given by
\begin{\eq}
\de_2 S
&=& \int d \eta d^3x \left\{ \f{2 a^2}{\ka^2} \left[ (1-3 \la) {\cal H} \left(
3 {\cal H} {\cal A}^2 + {\cal A} (2 \pa {\cal B}^i-h') \right)
+ (1-\la) (\pa_i {\cal B}^i)^2+\f{1}{2} \pa_i {\cal B}_j \pa^i {\cal B}^j \right. \right. \no \\
&&\left. \left. -\pa_i {\cal B}_j {h^{ij}}' +\f{1}{4} {h_{ij}}' {h^{ij}}'
+\la \left(\pa_i {\cal B}^i h'-\f{1}{4}h'^2 \right) \right]
+a^2 \xi \left({\cal A} +\f{1}{2} h \right)
\left( \pa_i \pa_j h^{ij} -\Delta  h \right) \right.  \no \\
&&\left.+a^2 \left[ \f{1}{2} \de \phi'^2-{\cal A} \phi_0' \de \phi' +\f{1}{2} {\cal A}^2 \phi_0'^2 +\pa_i {\cal B}^i \phi_0' \de \phi-\f{a^2}{2} V_{\phi_0 \phi_0} \de\phi^2 -a^2 V_{\phi_0} \de \phi {\cal A}-\f{1}{2} \phi_0' \de \phi h'\right] \right.\no \\
&&\left. -a^4 \left({\cal V}^{(2)}+\de Z \right)\right\},
\end{\eq}
where ${\cal V}^{(2)}$ is the quadratic part of the potential ${\cal V}$ and  $\Delta  \equiv \de^{ij} \pa_i \pa_j$, $\de Z=\sum_{n=1}^3 \xi_n~ \partial_i^{(n)}\de\phi \partial^{i(n)}\de\phi$.

Now, in order to separate the scalar, vector, and tensor contributions, we consider the most general (SVT) decompositions $(\pa_i S^i = \pa_i F^i = \tilde{H} = \pa_i \tilde{H}^i_j = 0 )$,
$
\calB_i =  \pa_i {\cal B} + S_i \, ,~
h_{ij} = 2 {\cal R} \de_{ij}+ \pa_i \pa_j {\cal E} + \pa_{(i} F_{j)} + \tilde{H}_{ij} \, .
$
Then, the pure tensor, vector, and scalar parts of the total action are given by, respectively,
\begin{\eq}
\label{reduced_2nd_tensor_action}
\delta_2 S^{(t)} &=& \int d \eta d^3x~ a^2
 \left[ \frac{2}{\kappa^2} {\tilde{H}_{ij}}' {\tilde H}^{ij'}
+\xi\tilde{H}_{ij}\Delta \tilde{H}^{ij}
+ \frac{\alpha_2}{a^2}\Delta \tilde{H}_{ij}\Delta \tilde{H}^{ij}+ \frac{\alpha_3}{a^3} \epsilon^{ijk} \Delta \tilde{H}_{il} \Delta  \pa_j \tilde{H}^l_{k}  \right. \no \\
&&~~~~~~~~~~~~~~~~~~~~\left.
 - \frac{\alpha_4}{a^4} \Delta \tilde{H}_{ij}\Delta^2 \tilde{H}^{ij} \right],\\
\label{reduced_2nd_vector_action}
\delta_2 S^{(v)} &=& \frac{1}{\kappa^2} \int  d \eta d^3x~ a^2 \pa_i
 \left( S^j-{F^j}' \right) \pa_i \left( S_j-F_j' \right) \, ,\\
\label{reduced_2nd_scalar_action}
\delta_2 S^{(s)}
&=&  \int d \eta d^3x ~a^2 \left\{ \frac{2(1-3\lambda)}{\kappa^2}
 \left[ 3{\calR'}^2 - 6\calH {\cal A}\calR' +
3\calH^2 {\cal A}^2
- 2\left( \calR' - \calH \cA \right) \Delta({\cal B}-{\cal E}')
\right] \right.
\nonumber\\
 &&+ \frac{2(1-\lambda)}{\kappa^2} \left[ \Delta\left({\cal B}-{\cal E}'\right) \right]^2
- 2\xi(\calR+2{\cal A})\Delta\calR
+ \frac{2}{a^2} \left( 8\alpha_1+3\alpha_2  \right)(\Delta\calR)^2
\nonumber\\
&& \left.
-\frac{2}{a^4} \left( 3\alpha_4+2\alpha_5+8 \al_6 \right)\Delta\calR \Delta^2 \calR -a^2V_{\phi_0} \cA\delta\phi
  -\frac{1}{2a^2}V_{\phi_0\phi_0}\delta\phi^2 - \delta{Z} \right.\no \\
&& \left.+ \frac{1}{2}{\delta\phi'}^2  - \phi_0'\delta\phi' {\cal A}
+\f{1}{2} {\phi_0'}^2 {\cal A}^2
+ [\Delta({\cal B}-{\cal E}')-3 \calR']\phi_0'\delta\phi \right\} \, .
\end{\eq}
Here, it is important to note that sixth-order-derivative terms in the gravity action contribute to scalar as well as tensor perturbations, through the specific combination of `$3\alpha_4+2\alpha_5+8 \al_6$' for the former but through only `$\al_4$' for the latter. This is what we need for renormalizability and scale-invariant spectrums (as we can see shortly), for both scalar and tensor. In the detailed-balanced case, the combination `$3\alpha_4+2\alpha_5+8 \al_6$' vanishes though $\al_4$ does not. So in that case, the theory would not be renormalizable and nor scale invariant for scalar part, which is in contradict to observational data !

Now, in order to exhibit the true dynamical degrees of freedom we consider the Hamiltonian reduction method \ci{Fadd}, for the cosmologically perturbed actions \ci{Garr,Gong} and, after some computation, the action reduces to
\begin{equation}
\label{uaction}
\delta_2S_\star^{(s)}
 = \int d\eta d^3x \frac{1}{2}\left\{ {u'}^2
 - u \left[ \f{1}{2}\left( \calG_1'\calG_1^{-1} \right)'
 - \f{1}{4}\left(\calG_1'\calG_1^{-1}\right)^2
 - \calG_1 \left( \calG_2 \calG_1^{-1} \right)'
 - \calG_2^2 + 4\calG_1\calG_3 \right]u \right\} \,
\end{equation}
for a true scalar degree of freedom $u$. In UV limit ($3 \tilde{\al_4} \equiv 3 \al_4 +2 \al_5+8 \al_6, ~z \equiv a \phi_0'/\cH$), its equations of motion reduce to
\begin{\eq}
\label{uequation_UV}
u'' =- \om_{u(UV)}^2 u  \, , ~~
\om_{u(UV)}^2=  \f{-6 \xi_3 \tilde{\al_4}}{a^2 z^2 } \left[ 2+\f{\ka^2 (1-\la)}{4 (1-3 \la)}\f{z^2}{a^2}\right] \De^3.
\end{\eq}
Here, it is important to note that there are sixth-spatial derivatives, as required by the scale invariance of the observed power spectrum \ci{Muko:1007} as well as the (power-counting) renormalizability \ci{Hora}. This occurs only when there are sixth-derivative terms in the starting scalar action  as well as some breaking of the detailed balance condition in sixth-derivative terms for the gravity action ({\it i.e.,} $\tilde{\al_4} \neq 0$)

Regarding the scale invariance of the power spectrums, it has been noted that
Ho\v{r}ava gravity could provide an alternative mechanism for the early Universe
without introducing the hypothetical inflationary epoch \ci{Muko:1007}: The basic
reason of the alternative mechanism comes from the momentum-dependent speeds of
gravitational perturbations which could be much larger than the current, low energy
({\it i.e., }IR) speed $c$ so that the exponentially expanding early space-time
could be mimicked. In order to see this explicitly in our case, we consider the
power-law expansions \ci{Lucc} ($t$ is the physical time, defined by $dt=a d \eta$),
$
a =a_0 t^p, ~(1/3 < p <1).
$
Then we can produce the scale-invariant power spectrums for
the quantum field $\hat{\zeta}$ of the $\zeta$ (scalar) perturbation, as follows,
\begin{\eq}
\left<0|
\hat{\zeta}_{\bf k} (\eta) \hat{\zeta}_{{\bf k}'} (\eta)
|0\right>
=(2 \pi)^3 \de ({\bf k}+{\bf k}') \f{2 \pi^2}{k^3} \De^2_{\zeta} (k),
~~\De^2_{\zeta}=\f{k^3}{2 \pi^2} | \zeta_{\bf k} |^2
=\f{\hbar}{8 \pi^2} \sqrt{\f{(3 \la-1)[1-{(1-\la)}/{4}] }{6 \ka^2 \xi_3 \tilde{\al}_4}}~,
\end{\eq}
without knowing the details of the history of the early
Universe and the form of the (non-derivative) potential $V (\phi)$.

This result shows that one can achieve the "inflation without inflation" picture, as argued by Mukohyama \cite{Muko:1007}. The basic reason of this is that, in Ho\v{r}ava gravity, due to the momentum-dependent, superluminal speeds of fluctuations is possible in the early Universe, which is assumed to be UV region, so that one can mimic "the inflationary scenario without inflationary epoch" ! (See Fig. 1 in \cite{Shin}.)

\section*{Acknowledgments}
This was supported by Basic Science Research Program through the National Research Foundation of Korea (NRF) funded by the Ministry of Education, Science and Technology {(2016R1A2B401304)}.

%
%

\end{document}